\documentclass[aps,pra,twocolumn,showpacs,showkeys,preprintnumbers,amsmath,amssymb,groupedaddress]{revtex4-1}

\usepackage{graphicx}
\usepackage{dcolumn}
\usepackage{bm}
\usepackage{CJK}
\begin{document}

\title{Modulational instability in nonlocal Kerr media with sine-oscillatory response}


\author{Zhuo Wang}
\author{Qi Guo}\email{guoq@scnu.edu.cn}
\author{Weiyi Hong}
\author{Wei Hu}
\affiliation{Guangdong Provincial Key Laboratory of Nanophotonic Functional Materials and Devices, South China Normal University, Guangzhou 510631, P. R. China}


\date{\today}

\begin{abstract}
We discuss the modulational instability (MI) of plane waves in nonlocal Kerr media with the sine-oscillation response function, which can model the nematic liquid crystal with negative dielectric anisotropy.
The results in the framework of the nonlocal
nonlinear Schr\"{o}dinger equation show that the MI in this system has two extraordinary properties. First, MI exists in both cases of negative Kerr coefficient $(n_{2}<0)$ and positive Kerr coefficient $(n_{2}>0)$. Second, the maximum gain of MI is not affected by the intensity of the plane wave, and the wave number 
of the maximum gain depends on the oscillation period of the response function. We also explore, for the first time to the best of our knowledge, the physical mechanism of MI in the framework of the four-wave mixing (FWM). 
By introducing a phase mismatch term $\triangle k$ and a (nonlinear) growth factor $\gamma$ (both of them have clear physical meanings), we find that the necessary and sufficient condition for the MI appearance is that the phase mismatching during the FWM process should be small enough such that $|\triangle k|<2|\gamma|$. Basing on the discover, we present a new scenario to uniformly and consistently explain why the MI can happen for the focusing local Kerr media and cannot for the defocusing local Kerr media, as well as the MI phenomena in the nonlocal cases.
\end{abstract}


\maketitle
\section{\label{sec-1}Introduction}
Modulational instability (MI) is a kind of ubiquitous instabilities in nonlinear systems. It signifies the amplification of 
random perturbations on a harmonic wave 
when it propagates in nonlinear media. The growth of the perturbations generates spectral sidebands, which makes the harmonic wave evolves into a modulated state. The phenomena of MI have been studied and identified in various physical systems, such as fluids~\cite{Benjamin-jfm-67}, plasma~\cite{Hasegawa-pra-70} and nonlinear optics~\cite{Bespalov-jeptl-67,Ostrovskii-sp-67,Hasegawa-ol-84,Tai-prl-86,Agrawal-book}, etc. In the context of optical fiber, MI generally lead to the breakup of a continuous wave into a periodic pulse train\cite{Hasegawa-ol-84,Tai-prl-86}. And for a monochromatic spatial optical beam, the effect of MI is to split the beam into a transversely periodic array of fine localized structures (filamentation)\cite{Bespalov-jeptl-67}. Thus, in optical Kerr media, MI is seen as a precursor for the formation of bright solitions, whereas dark solitons requires the absence of MI.

Spatial nonlocality is one of the most important properties
of the optical Kerr effect. The nonlocality means that the light-induced nonlinear refractive index (NRI) at a given point is determined not only by the light intensity at that point but also by the light intensity near that point, which can be described phenomenologically as~\cite{Guo-book}
\begin{equation}\label{equ-0}
\Delta n=n_{2}\int^{\infty}_{-\infty}R(\bm{r}_{\perp}-\bm{r}'_{\perp})|E(\bm{r}'_{\perp},z)|^{2}d\bm{r}'_{\perp}, \end{equation}
where $\Delta n$ is the NRI, $n_{2}$ is the Kerr coefficient determined by material properties, $\bm{r}_{\perp}$ is the transverse coordinate vector, the real symmetric function $R(\bm{r}_{\perp})$ is the response function of the nonlocal optical Kerr media, and $E$ is the optical field. Since Snyder and Mitchell raised the attention to nonlocality~\cite{Snyder-sci-97}, nonlocal spatial optical solitons have been systematically researched in the optical Kerr media with strong nonlocality, including 
nematic liquid crystals (NLC) (with positive dielectric anisotropy)~\cite{Conti-prl-03,Conti-prl-04,Rasmussen-pre-05,Hu-apl-06,Assanto-book,Assanto-book-2} and lead glasses~\cite{Rotschild-prl-05,Rotschild-NaturePhys-06,Lu-pra-08,Shou-ol-09,Shou-ol-11}, etc. Recently, bright optical solitons were observed in the planar cell containing the NLC with negative dielectric anisotropy~\cite{Wang-arxiv-14}. In the 1+1 dimensional model of this system, the NRI is described by~\cite{Wang-arxiv-14,Liang-arxiv-15}
\begin{equation}\label{equ-1}
w_{m}^{2}\frac{d^{2}\Delta n}{dx^{2}}+\Delta n=n_{2}|E|^{2},
\end{equation}
where 
$w_{m}$ is a positive constant representing the nonlinear characteristic length, and the Kerr coefficient $n_{2}$ is negative. The response function derived from Eq.~(\ref{equ-1}) is sine-oscillatory
\begin{equation}\label{equ-2}
R(x)=\frac{1}{2w_{m}}\sin\left(\frac{|x|}{w_{m}}\right).
\end{equation}
This response function was first obtained in the model of quadratic solitons~\cite{Nikolov-pre-03}, although the nonlinear process of quadratic solitons is the second-order nonlinear effect rather than the third-order one.
Both of bright solitons and dark solitons have been found in the medium modelled by Eq.~(\ref{equ-1})~\cite{Liang-arxiv-15},
but the dark solitons are unstable. Thus discussing the MI in the spatially nonlocal optical Kerr medium with the sine-oscillatory response function helps to understand the stability of the bright solitons and the instability of the dark solitons in such a system.

For spatially nonlocal nonlinear systems, the nonlocality has strong influences on the MI. 
 The first theoretical result about the MI in nonlocal optical Kerr media~\cite{Krolikowski-pre-01,Wyller-pre-02} showed that the gain spectrum of the MI is affected by the characteristic length of the nonlocality and the profile of the response function.
 Then, Assanto's group observed the 
MI in the NLC with positive dielectric anisotropy~\cite{Peccianti-pre-03}, which is the first experimental evidence of the MI in spatially nonlocal optical Kerr media. Later, Wyller~{\em et~al.} discussed the MI in the model of quadratic solitons with the sine-oscillatory response~\cite{Wyller-pd-07}. In Wyller's research, the effective NRI, which is the second harmonic optical field,
is induced by the square of the fundamental wave $E^{2}$, while in spatially nonlocal optical Kerr media the NRI is induced by the light intensity $|E|^{2}$ as shown in Eq.~(\ref{equ-0}). Therefore, the property of the MI in spatially nonlocal optical Kerr media with the sine-oscillatory response function can be different from that in the model of quadratic solitons.

In this paper, we analytically study the MI in the spatially nonlocal optical Kerr medium with the sine-oscillatory response, and obtain a insight into the physical mechanism of MI with the theory of four-wave mixing (FWM). The contents are organized as follows. In Sec.~\ref{sec-2}, within the framework of the nonlocal nonlinear Schr\"{o}dinger equation~\cite{Krolikowski-pre-01}, we discuss the MI 
in the nonlocal optical Kerr system with the sine-oscillatory response function, and find that the MI in this system has two extraordinary properties.
The aim of Sec.~\ref{sec-3} is to discuss the physical mechanism of MI in the framework of the FWM interaction. 
\section{\label{sec-2}MI in Kerr media with sine-oscillatory response}
In the 1+1 dimensional lossless nonlocal optical Kerr media, 
a linearly-polarized monochromatic optical beam propagating along the $z$ axis can be described by the nonlocal nonlinear Schr$\ddot{\textrm{o}}$dinger equation (NNLSE)~\cite{Krolikowski-pre-01,Guo-book,Assanto-book-2}
\begin{equation}\label{equ-3}
i\frac{\partial U}{\partial z}+\frac{1}{2k}\frac{\partial^{2} U}{\partial x^{2}}+\frac{n_{2}k}{n_{0}}U\int_{-\infty}^{\infty}R(x-x')|U(x',z)|^{2}dx'=0,
\end{equation}
where $U$ is the slowly varying complex amplitude of the electric field satisfying the relation that $E=U\exp(ikz)$, $k$ is the wave number in the medium with a linear refractive index $n_{0}$, and the Kerr coefficient $n_{2}$ could be either positive or negative. For different nonlocal optical Kerr systems, the response functions $R(x)$ are different, such as the Gaussian function~\cite{Krolikowski-pre-01}, the exponential-decay function~\cite{Rasmussen-pre-05}, the rectangular function~\cite{Krolikowski-pre-01} and the sine-oscillatory function~\cite{Wang-arxiv-14,Liang-arxiv-15}, etc.

The NNLSE has a plane wave solution
\begin{equation}\label{equ-4}
\overline{U}=\sqrt{I_{0}}\exp\left[i\frac{2\pi n_{2}kI_{0}\widetilde{R}(0)}{n_{0}}z\right],
\end{equation}
where $I_{0}$ is its intensity, and $\widetilde{R}(k_{x})$ represents the Fourier transform of the response function
\begin{equation}\label{equ-5}
\widetilde{R}(k_{x})=\frac{1}{2\pi}\int_{-\infty}^{\infty}R(x)\exp(-ik_{x}x)dx.
\end{equation}
Then we add a random perturbation on the plane wave such that
\begin{equation}\label{equ-6}
U(x,z)=\left[\sqrt{I_{0}}+\psi(x,z)\right]\exp\left[i\frac{2\pi n_{2}kI_{0}\widetilde{R}(0)}{n_{0}}z\right],
\end{equation}
 and examine evolution of the perturbation $\psi(x,z)$ using a linear stability analysis. Substituting Eq.~(\ref{equ-6}) into the model~(\ref{equ-3}) and linearizing with respect to $\psi$ as $|\psi (x,z)|^{2}\ll I_{0}$, we obtain the linearized evolution equation of the perturbation
\begin{equation}\label{equ-7}
i\frac{\partial\psi}{\partial z}+\frac{1}{2k}\frac{\partial^{2}\psi}{\partial x^{2}}+\frac{2n_{2}kI_{0}}{n_{0}}\int_{-\infty}^{\infty}R(x-x')\text{Re}[\psi(x',z)]dx'=0.
\end{equation}
Decomposing the perturbation into real and imaginary parts $\psi=u+iv$, and using the Fourier transform shown in Eq.~(\ref{equ-5}), we derive a set of ordinary differential equations in $k_{x}$ domain from Eq.~(\ref{equ-7})
\begin{subequations}\label{equ-8}
\begin{equation}
\frac{d^{2} \widetilde{u}}{d z^{2}}+k_{x}^{2}\left[\frac{k_{x}^{2}}{4k^{2}}-\frac{2\pi n_{2}I_{0}\widetilde{R}(k_{x})}{n_{0}} \right]\widetilde{u}=0,\label{equ-8a}
\end{equation}
\begin{equation}
\frac{d^{2} \widetilde{v}}{d z^{2}}+k_{x}^{2}\left[\frac{k_{x}^{2}}{4k^{2}}-\frac{2\pi n_{2}I_{0}\widetilde{R}(k_{x})}{n_{0}} \right]\widetilde{v}=0.\label{equ-8b}
\end{equation}
\end{subequations}
By solving Eqs.~(\ref{equ-8}), we can obtain
\begin{equation}\label{equ-11}
\widetilde{\psi}(k_{x},z)=\widetilde{u}+i\widetilde{v}
=c_{1}\exp(\lambda z)+c_{2}\exp(-\lambda z),
\end{equation}
%
where $c_{1}$ and $c_{2}$ are arbitrary constants, and the eigenvalue $\lambda$ is given by
\begin{equation}\label{equ-10}
\lambda=|k_{x}|\sqrt{\frac{2\pi n_{2}I_{0}\widetilde{R}(k_{x})}{n_{0}}-\frac{k_{x}^{2}}{4k^{2}}}.
\end{equation}

Notice that if $\lambda$ is real, the spatial spectrum component of the perturbation $\widetilde{\psi}$ grows exponentially with $z$, so the nonlinear system has MI. By contrast, when $\lambda$ is imaginary the plane wave is stable.


Now, we consider the case that the nonlocal response function is of the sine-oscillatory type given by Eq.~(\ref{equ-2}), and the Fourier transform of the response function is~\cite{Wyller-pd-07}
\begin{equation}\label{equ-12}
\widetilde{R}(k_{x})=\frac{1}{2\pi (1-w_{m}^{2}k_{x}^{2})}.
\end{equation}
Thus the  gain coefficient that is defined by $\text{g}=2\text{Re}\left[\lambda\right]$ can be obtained as
\begin{equation}\label{equ-14}
\text{g}=\text{Re}\left[2|k_{x}|\sqrt{\frac{n_{2}I_{0}}{n_{0}(1-w_{m}^{2}k_{x}^{2})}-\frac{k_{x}^{2}}{4k^{2}}}\right],
\end{equation}
which only exists when $n_{2}I_{0}/n_{0}(1-w_{m}^{2}k_{x}^{2})>k_{x}^{2}/4k^{2}$. As the gain coefficient is an even function of $k_{x}$, we only discuss the variation of $\text{g}$ with $|k_{x}|$.

\begin{figure}[htb]
\includegraphics[width=8cm]{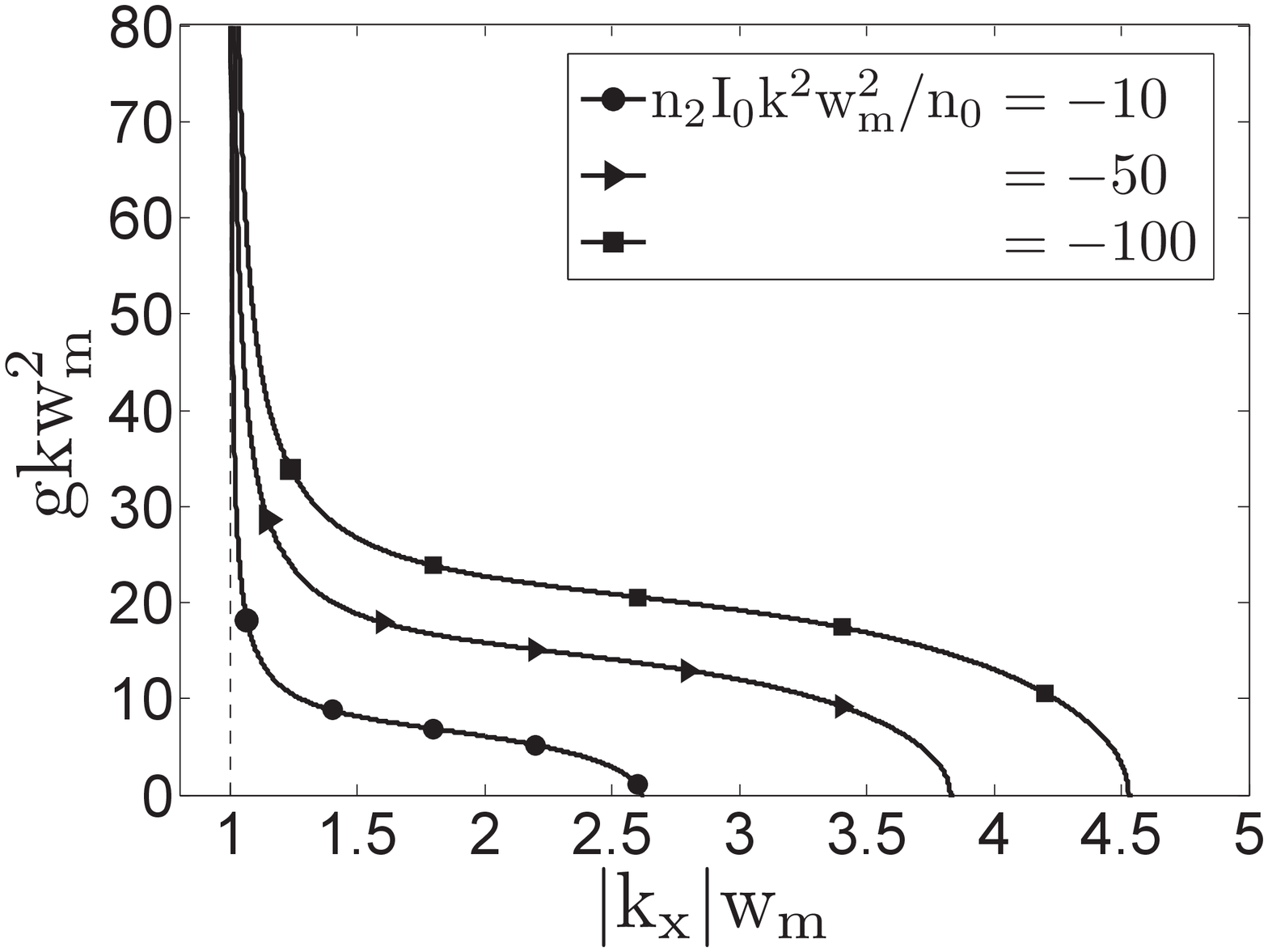}
\caption{\label{Fig1} The gain spectra of MI with different values of $I_{0}$ in the case $n_{2}<0$.}
\end{figure}
In the case $n_{2}<0$, the MI occurs when
\begin{equation}\label{equ-15}
1<w_{m}^{2}k_{x}^{2}<\frac{1}{2}\sqrt{1-\frac{16w_{m}^{2}k^{2}n_{2}I_{0}}{n_{0}}}+\frac{1}{2}.
\end{equation}
Fig.~\ref{Fig1} shows the gain spectra of MI with three values of the intensity of the plane wave. 
We can see that when the light intensity $I_{0}$ increases, the region of MI expands and the gain coefficient at the same wave number point also increases. So the intensity of the plane wave tends to boost MI in this case. And an interesting characteristic of the gain spectra is that the maximum gain points do not move with the intensity of the plane wave. Instead, they exist at the points of $|k_{x}|=1/w_{m}$ fixedly. In addition, because the Kerr coefficient is negative $(n_{2}<0)$, the MI bands only appear in the region where the spectral function~(\ref{equ-12}) has negative values.

\begin{figure}[htb]
\includegraphics[width=8cm]{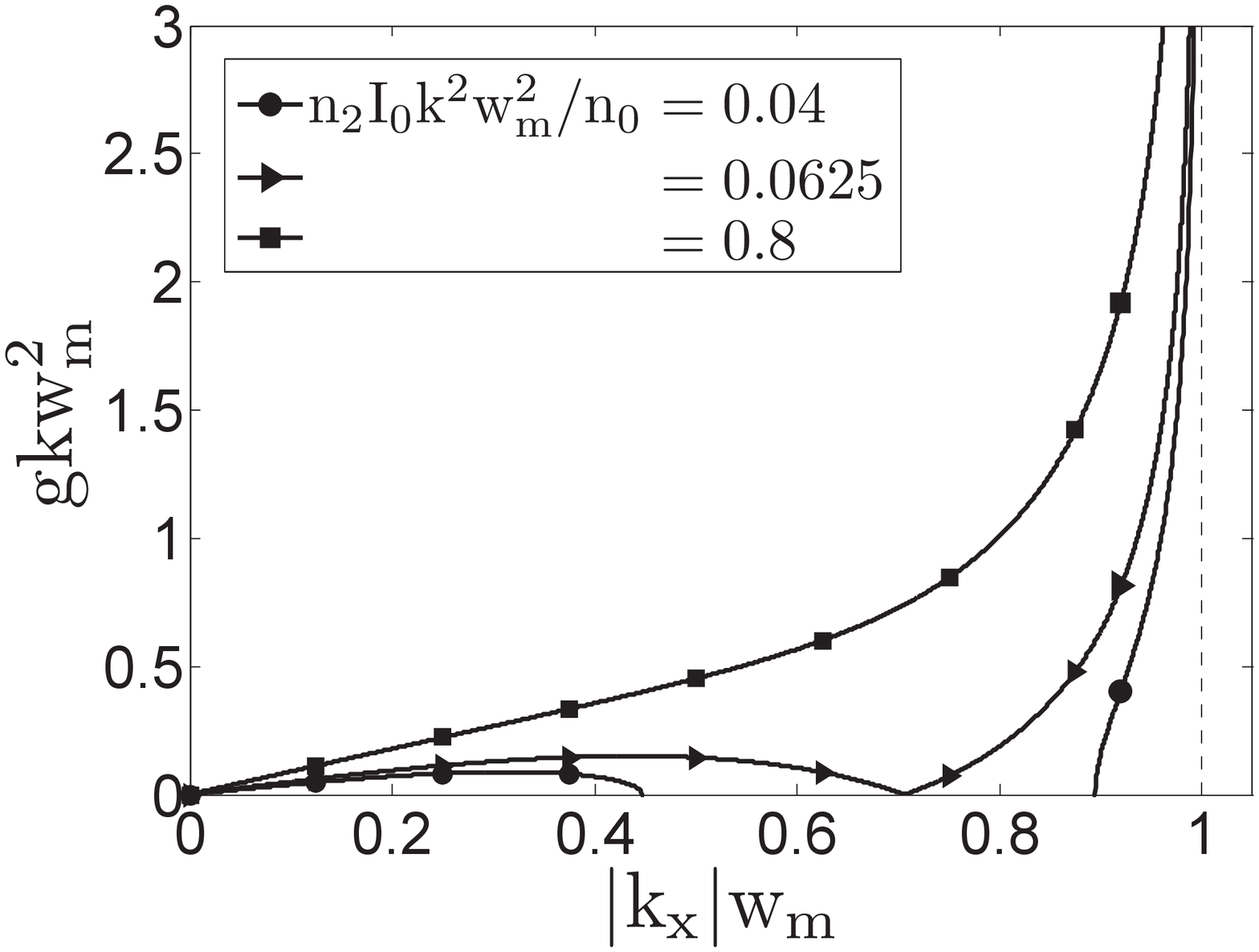}
\caption{\label{Fig2} The gain spectra of MI with different values of $I_{0}$ in the case $n_{2}>0$.}
\end{figure}
In the case $n_{2}>0$, MI also can happen, 
but the number of the MI gain bands is affected by the intensity of the plane wave, as shown in Fig~\ref{Fig2}. When the light intensity 
is small enough to meet the condition
\begin{equation}\label{equ-16}
0<I_{0} \leq \frac{n_{0}}{16w_{m}^{2}k^{2}n_{2}},
\end{equation}
there are two MI gain bands in the each side of the origin
\begin{equation}\label{equ-17}
\left\{
\begin{array}{c}
0<w_{m}^{2}k_{x}^{2}<\frac{1}{2}-\frac{1}{2}\sqrt{1-\frac{16w_{m}^{2}k^{2}n_{2}I_{0}}{n_{0}}},
\\
\\
1>w_{m}^{2}k_{x}^{2}>\frac{1}{2}+\frac{1}{2}\sqrt{1-\frac{16w_{m}^{2}k^{2}n_{2}I_{0}}{n_{0}}}.
\end{array}\right.
\end{equation}
And after $I_{0}$ exceeds the critical value $n_{0}/16w_{m}^{2}k^{2}n_{2}$, the MI gain bands combine into one
\begin{equation}\label{equ-18}
0<w_{m}^{2}k_{x}^{2}<1.
\end{equation}
Again, MI is being boosted by the light intensity $I_{0}$, and the maximum MI gain also appears at the points of $|k_{x}|=1/w_{m}$ unchangeably. However, different from the case $n_{2}<0$, the MI gain in this case appears in the region where the spectral function~(\ref{equ-12}) has positive values.

By summarizing the results above, MI can occur in both cases of negative Kerr coefficient $(n_{2}<0)$ and positive Kerr coefficient $(n_{2}>0)$ in the optical Kerr media with the sine-oscillatory response. Moreover, we find that MI only happens when the spectrum function~(\ref{equ-12}) has the same sign with the Kerr coefficient $n_{2}$. As a matter of fact, $n_{2}\widetilde{R}>0$ is a necessary condition for the existence of MI in Kerr media, which will be illustrated in the next section. This conclusion can also explain the well-known results that~\cite{Wyller-pre-02} MI only appears when $n_{2}>0$ for the response functions with a positive-definite spectrum, such as the Gaussian function and the exponential-decay function, as well as that~\cite{Krolikowski-pre-01}
the rectangular response function which have either negative or positive spectral values can lead to MI in both cases that $n_{2}<0$ and $n_{2}>0$. 
So far, only two response functions, the rectangular function and the sine-oscillatory function, have been found to have such a property that the MI can occur in both cases that $n_{2}<0$ and $n_{2}>0$, but the former is only a phenomenological one that can not describe any real physical systems. The sine-oscillatory response, however, can describe the nonlinear response in the NLC with negative dielectric anisotropy. Thus the occurrence of MI with the negative Kerr coefficient $(n_{2}<0)$ for the sine-oscillatory response can support the facts of the stability of the bright solitons
~\cite{Wang-arxiv-14,Liang-arxiv-15} and the instability of the dark solitons~\cite{Liang-arxiv-15} in the NLC with negative dielectric anisotropy. Furthermore, the largest gain of the MI fixedly occurs at the points of $|k_{x}|=1/w_{m}$ in the system discussed here, 
while the maximum MI gain points move with the light intensity in the Kerr systems with other response functions~\cite{Wyller-pre-02} or in the system of quadratic solitons~\cite{Wyller-pd-07}, even though the latter also has the sine-oscillatory response. 
In the system discussed here, since the transverse wave numbers of the maximum gain are independent of the light intensity,  the repetition period 
of the periodic pattern of the MI would not change with the light intensity, and this feature may be demonstrated through experiments in the NLC with negative dielectric anisotropy. 

\section{\label{sec-3}The physical mechanism of MI: phase mismatch vs growth factor}
The optical Kerr effect is a kind of third-order nonlinear effects. The propagation of an optical beam in an optical Kerr medium is influenced by the third-order nonlinear polarization. 
Another way, therefore, to discuss the MI is analyzing the interaction of the strong plane wave and the perturbations through the third-order nonlinear polarization by coupled-wave equations~\cite{Agrawal-book,Cappellini-pra-91,book-Boyd}. In this section, we will discuss the MI in the framework of the four-wave mixing (FWM) interaction. We will introduce a phase mismatch term and a (nonlinear) growth factor, both of which have clear physical meanings, and show how the competition between them trigger off or restrain the MI
in the optical Kerr media.

\subsection{\label{subsec-3a}Evolution equations for perturbations}
The paraxial wave equation of a linearly-polarized monochromatic optical beam in nonlocal optical Kerr media can also be re-expressed as~\cite{book-Boyd}
\begin{equation}\label{equ-18a}
2ik\frac{\partial U}{\partial z}+\frac{\partial^{2} U}{\partial x^{2}}=-\frac{k^{2}}{\epsilon_{0}n_{0}^{2}}P^{\text{NL}}_{\text{nonlc}},
\end{equation}
where $\epsilon_{0}$ is the dielectric constant of vacuum, and $P^{\text{NL}}_{\text{nonlc}}$ is the third-order nonlinear polarization in nonlocal optical Kerr media. Through the analogy of the relation between the third-order nonlinear polarization and the NRI in local optical Kerr media, we get a phenomenological description of the nonlinear polarization in nonlocal optical Kerr media (see the Appendix A for details)
\begin{equation}\label{equ-18b}
P^{\text{NL}}_{\text{nonlc}}=2n_{0}\epsilon_{0}n_{2}U\int^{\infty}_{-\infty}R(x-x')|U(x',z)|^{2}dx'.
\end{equation}
We regard the strong plane wave as a pump wave, and consider that the perturbation on the plane wave consists of signal waves which have different transverse wave numbers. So it is convenient to assume that
\begin{equation}\label{equ-19}
U(x,z)=A_{0}(z)+\eta(x,z)\ \ \ (|\eta|\ll |A_{0}|),
\end{equation}
where $A_{0}(z)$ is the complex amplitude of the pump wave, 
and $\eta(x,z)$ is the complex amplitude of the perturbation. In the preceding section, the discussion about MI is based on the linear stability analysis, which only describes the initial evolution of the perturbation (when the condition of $|\eta|\ll |A_{0}|$ is satisfied). 
Such a discussion ignores the contribution from the higher order terms (more than one order) of the perturbation to the nonlinear polarization. It means that we neglect the pump depletion and only consider the nonlinear coupling of a pair of the signal waves which have axisymmetric wave vectors respect to the $z$ axis. Therefore, we can treat the evolution of random perturbations in a simplified way that assuming the perturbation only consists of a pair of symmetrically displaced signal waves, and the total optical field is
\begin{equation}\label{equ-21}
U(x,z)=A_{0}(z)+A_{+}(z)\exp(iqx)+A_{-}(z)\exp(-iqx).
\end{equation}
where $|A_{+}|\sim |A_{-}|\ll |A_{0}|$, and $\pm q$ are transverse wave numbers. Following the procedure presented in Ref.~\cite{book-Boyd} and by substituting Eq.~(\ref{equ-21}) into Eq.~(\ref{equ-18b}), we can obtain
\begin{equation}\label{equ-22}
P^{\text{NL}}_{\text{nonlc}}=P_{0}(z)+P_{+}(z)\exp(iqx)+P_{-}(z)\exp(-iqx),
\end{equation}
where $P_{0}$, $P_{+}$ and $P_{-}$ are nonlinear polarizations that are, respectively, phase matched to $A_{0}$, $A_{+}$ and $A_{-}$ along the $x$ direction, and
\begin{subequations}\label{equ-23}
\begin{equation}
P_{0}=4\pi n_{0}\epsilon_{0}n_{2}\widetilde{R}(0)|A_{0}|^{2}A_{0},\label{equ-23a}
\end{equation}
\begin{eqnarray}
\nonumber
P_{\pm}=4\pi n_{0}\epsilon_{0}n_{2}&&\left[\widetilde{R}(0)|A_{0}|^{2}A_{\pm}+\widetilde{R}(q)|A_{0}|^{2}A_{\pm}\right. \\
&&\left.+\widetilde{R}(q)A^{2}_{0}A^{*}_{\mp}\right].\label{equ-23b}
\end{eqnarray}
\end{subequations}
In the procedure above, we have neglected the higher order terms of $A_{\pm}$ and used the spectral feature of a real symmetric function $R(x)$ that $\widetilde{R}(q)=\widetilde{R}^{*}(q)=\widetilde{R}(-q)$. Then inserting Eqs.~(\ref{equ-21}), (\ref{equ-22}) and (\ref{equ-23}) into Eq.~(\ref{equ-18a}), we get the coupled-wave equations to describe the FWM interaction of the pump wave and the signal waves
\begin{subequations}\label{equ-24}
\begin{equation}
\frac{\text{d} A_{0}}{\text{d} z}=i\frac{2\pi n_{2}k}{n_{0}}\widetilde{R}(0)|A_{0}|^{2}A_{0},\label{equ-24a}
\end{equation}
\begin{eqnarray}
\nonumber
\frac{\text{d} A_{\pm}}{\text{d} z}&&=-i\frac{q^{2}}{2k}A_{\pm}+i\frac{2\pi n_{2}k}{n_{0}}\left[\widetilde{R}(0)|A_{0}|^{2}A_{\pm}\right. \\
&&\left.+\widetilde{R}(q)|A_{0}|^{2}A_{\pm}+\widetilde{R}(q)A^{2}_{0}A^{*}_{\mp}\right].\label{equ-24b}
\end{eqnarray}
\end{subequations}
Eq.~(\ref{equ-24a}) tells us that the pump wave is only phase-modulated by itself, and its expression is
\begin{equation}\label{equ-25}
A_{0}=\sqrt{I_{0}}\exp(ik_{0z}^{NL}z),
\end{equation}
where $k_{0z}^{NL}=2\pi n_{2}\widetilde{R}(0)I_{0}k/n_{0}$ is the additional wave number along the $z$ axis contributed by self-phase modulation (SPM). Substituting Eq.~(\ref{equ-25}) into Eqs.~(\ref{equ-24b}), we obtain that
\begin{equation}\label{equ-26}
\frac{\textrm{d} A_{\pm}}{\textrm{d} z}=-i\frac{q^{2}}{2k}A_{\pm}+ik_{\pm z}^{NL}A_{\pm}+i\gamma A_{\mp}^{*}\exp(i2k_{0z}^{NL}z).
\end{equation}
where
\begin{equation}\label{equ-26a}
k_{+z}^{NL}=k_{-z}^{NL}=\frac{2\pi n_{2}I_{0}k}{n_{0}}[\widetilde{R}(0)+\widetilde{R}(q)],
\end{equation}
\begin{equation}\label{equ-26b}
\gamma (q)=\frac{2\pi n_{2}I_{0}k\widetilde{R}(q)}{n_{0}}.
\end{equation}
On the right-hand side of Eqs.~(\ref{equ-26}), the first term represents the phase delaying by $-q^{2}/2k$ along the $z$ axis resulting from the off-axis wave vectors. The second term represents the cross-phase modulation (XPM) from the pump wave 
that adds additional wave numbers to the signal waves along the $z$ axis. The last term is responsible for the energy transfer from the pump wave to the signal waves~\cite{book-Shen}. Thus $\gamma$ is named growth factor which relates to the growth rate of the signal waves, as will be discussed later. For further discussion, we substitute the expression
$$A_{\pm}(z)=a_{\pm}(z)\exp\left[ik_{\pm z}^{NL}z-i\frac{q^{2}}{2k}z\right]$$
into Eqs.~(\ref{equ-26}), and obtain the evolution equations for the signal waves
\begin{equation}\label{equ-27}
\frac{\textrm{d}^{2}a_{\pm}}{\textrm{d}z^{2}}-i\Delta k\frac{\textrm{d}a_{\pm}}{\textrm{d}z}-\gamma^{2}a_{\pm}=0,
\end{equation}
where
\begin{equation}\label{equ-28}
\Delta k=\frac{q^{2}}{k}-2\gamma.
\end{equation}

\subsection{\label{subsec-3b}Physical effects of $\Delta k$-term and $\gamma$-term}
Eqs.~(\ref{equ-27}) are the evolution equations for the perturbations (the signal waves). We should notice that the $\gamma$ in Eq.~(\ref{equ-28}) does not contribute at all to the gain of the signal waves, which will be discussed later, although it is same quantitatively with the growth factor $\gamma$ in the third term of Eqs.~(\ref{equ-27}).

In order to understand the physical effects of $\Delta k$-term and $\gamma$-term in Eqs.~(\ref{equ-27}), we first
should know what is $\Delta k$ physically. For this purpose, we consider the full expressions of the pump wave and the signal waves, which are
$E_{0}=\sqrt{I_{0}}\exp(i\bm{k}_{0}\cdot\bm{r}-i\omega t)$,
$E_{\pm}=a_{\pm}\exp(i\bm{k}_{\pm}\cdot\bm{r}-i\omega t)$,
where
$\bm{k}_{0}=(k+k_{0z}^{NL})\bm{e}_{z}$ and
$\bm{k}_{\pm}=(k-\frac{q^{2}}{2k}+k_{\pm z}^{NL})\bm{e}_{z}\pm q\bm{e}_{x}$
are the total wave vectors contributed by the linear refractive index and the nonlinear self- or cross-phase modulations, $\bm{r}=x\bm{e}_{x}+y\bm{e}_{y}+z\bm{e}_{z}$ ($\bm{e}_{x}$, $\bm{e}_{y}$, and $\bm{e}_{z}$ are, respectively, the unit vector along the $x$ axis, $y$ axis and $z$ axis). 
The phase mismatches of $E_{+}$ and $E_{-}$ with their corresponding cross-coupled polarizations $E_{0}^{2}E_{-}^{*}$ and $E_{0}^{2}E_{+}^{*}$ are~\cite{book-Shen}, respectively, $\Delta\bm{k}_+=2\bm{k}_{0}-\bm{k}_{-}-\bm{k}_{+}$
 and $\Delta\bm{k}_-=2\bm{k}_{0}-\bm{k}_{+}-\bm{k}_{-}$
, and have the same value
%
\begin{equation}\label{equ-29}
\Delta\bm{k}_+=\Delta\bm{k}_-=
\Delta k\bm{e}_{z}.
\end{equation}
The phase mismatches have only the component of the $z$ axis because the condition of the phase matching along the $x$ axis is used in the process to obtain the evolution equations above. It is clear now that $\Delta k$ represents the value of the phase mismatch between the signal waves and their corresponding cross-coupled nonlinear polarizations. The phase mismatch given by Eq.~(\ref{equ-29}) is the same in definition with that in the process of the new frequency generations, such as the second harmonic one and the third harmonic one, and so on~\cite{book-Boyd2,book-Shen2}. 
The difference, however, is that the wave vectors in the process of the new frequency generations are only contributed by the linear refractive index, while the wave vectors in our discussion here are contributed by both the linear refractive index and the NRI from SPM or XPM because the NRI must be taken into consideration 
in the self-acting nonlinear process.
As an example, we show the wave vector relation between the signal wave $E_{+}$ and the cross-coupled nonlinear polarization $E_{0}^{2}E_{-}^{*}$ in Fig.~\ref{fig3}.
\begin{figure}[htb]
\includegraphics[width=8cm]{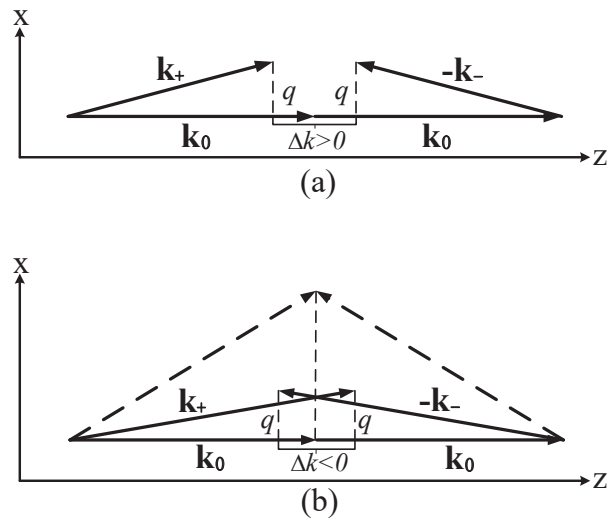}
\caption{\label{Fig3}The wave vector relation between $E_{+}$ and $E_{0}^{2}E_{-}^{*}$. (a): the case when $|\bm{k}_{\pm}|\leq|\bm{k}_{0}|$, (b): the case that $|\bm{k}_{\pm}|>|\bm{k}_{0}|$. The scale of the transverse wave number $q$ has been exaggerated in order to show the geometrical relationship clearly. In fact, the condition that $q\ll k$ should be satisfied under the paraxial approximation.}
\end{figure}
The geometry in Fig.~\ref{Fig3}(a) is the case that $|\bm{k}_{\pm}|\leq|\bm{k}_{0}|$, these four arrows can not form a triangle with any value of $q$, so phase match will not occur in this case. In Fig.~\ref{Fig3}(b) is the case that $|\bm{k}_{\pm}|>|\bm{k}_{0}|$. The solid arrows show the situation when $q$ is small, and $|\bm{k}_{\pm}|$ is too large to satisfied the condition of phase match. Phase matching, however, can be realized when the transverse wave number $q$ is large enough as shown by the dashed arrows. According to the analysis above, phase matching only can be achieved when $|\bm{k}_{\pm}|>|\bm{k}_{0}|$. This indicates that the effect of XPM on the signal waves needs to be larger than the effect of SPM on the pump wave (weak-wave retardation), which was first pointed out by Chiao~{\em et~al.} for the local optical Kerr effects~\cite{Chiao-prl-66}.

Now we can discuss the physical effects of the phase mismatch term ($\Delta k$-term) and the growth factor term ($\gamma$-term) in Eqs.~(\ref{equ-27}). To do this, we treat Eqs.~(\ref{equ-27})
in the way as if the $\Delta k$-term and the $\gamma$-term played separately an alone role in the evolution of the signal waves.
First, we consider the $\Delta k$-term alone and obtain the equation
\begin{equation}\label{equ-30}
\frac{\textrm{d}^{2}a_{\pm}}{\textrm{d}z^{2}}-i\Delta k\frac{\textrm{d}a_{\pm}}{\textrm{d}z}=0,
\end{equation}
and its solution is
\begin{equation}\label{equ-31}
a_{\pm}(z)=c_{3}+c_{4}\exp(i\Delta\textrm{k}z),
\end{equation}where $c_3$ and $c_4$ are integral constants. The result
shows that the function of the $\Delta k$-term is to fix
the signal wave amplitude to 
a certain constant and 
to forbid it to be amplified by the pump.
Second, we only consider the effect of the $\gamma$-term by setting $\Delta k=0$ in Eqs.~(\ref{equ-27}) and obtain
\begin{equation}\label{equ-32}
\frac{\textrm{d}^{2}a_{\pm}}{\textrm{d}z^{2}}-\gamma^{2}a_{\pm}=0.
\end{equation}
One of the two characteristic solutions 
of Eqs.~(\ref{equ-32}) is
\begin{equation}\label{equ-33}
a_{\pm}\sim \exp\left(|\gamma| z\right).
\end{equation}
It indicates that the effect of the $\gamma$-term is to amplify the signal waves and let them grow exponentially with the propagation distance.

The $\Delta k$-term and the $\gamma$-term, however, can not play a role separately in fact, and the two opposite effects will compete with each other to together determine the evolution of the signal waves. Such a competing process can be observed clearly by the following equation
\begin{equation}\label{equ-34}
\frac{d^{2}f_{\pm}}{dz^{2}}-\lambda^{2}f_{\pm}=0
\end{equation}
with
\begin{equation}\label{equ-35}
\lambda=\sqrt{\gamma^{2}-\frac{\Delta k^{2}}{4}},
\end{equation}which is obtained by substituting the transform $a_{\pm}=f_{\pm}\exp(iz\Delta k/2)$ into Eqs.~(\ref{equ-27}).
From Eqs.~(\ref{equ-34}) plus (\ref{equ-35}), we can see that the phase mismatch $\Delta k$ will suppress the growth of the signal waves. If $\Delta k$ is large enough such that the eigenvalue $\lambda$ becomes imaginary, $f_\pm$ will be harmonic oscillation, which means the energy transfer from the pump to the signals is completely inhibited. 

Comparing with the preceding result, Eq.~(\ref{equ-35}) together with Eq.~(\ref{equ-28}) coincides with Eq.~(\ref{equ-10}), and this indicates the equivalency between the framework of the FWM interaction and that of the NNLSE,
as suggested in Ref.~\cite{book-Boyd,Cappellini-pra-91,Agrawal-book}.

\subsection{\label{subsec-3c}Necessary and sufficient condition for MI}
As discussed above, the MI happens when the eigenvalue $\lambda$ is real. Therefore we can obtain by Eq.~(\ref{equ-35}) the necessary and sufficient condition for the MI
\begin{equation}\label{equ-36a}
|\Delta k|<2|\gamma|.
\end{equation}
This means that the MI must happen as long as the phase mismatch $\Delta k$ is small enough such that the condition above is satisfied. On the contrary, the transfer of energy from the strong plane wave to the perturbations will stop when $\Delta k$ is large enough, and the MI can never occur. In the case that $n_2$ and $\widetilde{R}$ are opposite in sign such that $\gamma<0$, Eq.~(\ref{equ-28}) tells us that Eq.~(\ref{equ-36a}) can not be satisfied~\cite{footnote}, thus MI never happens in this case, which is that shown in Fig.~\ref{Fig3}(a).
On the other hand, if $n_2$ and $\widetilde{R}$ have the same sign, we have $\gamma>0$.
As a result,
the substitution of Eq.~(\ref{equ-28}) into (\ref{equ-36a}) gives the spectral range for the MI~\cite{footnote}
\begin{equation}\label{equ-36b}
0<|k_{x}|<2\sqrt{\gamma k},
\end{equation}
where $q$ is replaced by $k_{x}$. This dose be the case shown in Fig.~\ref{Fig3}(b).

This necessary and sufficient condition and its deduction can explain all of the well-known results. For the local case, 
$\widetilde{R}\equiv1/2\pi$, thus the growth factor $\gamma$ is a constant.
When $n_{2}>0$, $\gamma$ is positive, we therefore always have a certain 
$k_x$-band that is determined by Eq.~(\ref{equ-36b}), 
and MI must happen in this band.
In contrast, $\gamma$ is negative when $n_{2}<0$, thus the plane wave solution is stable~\cite{Bespalov-jeptl-67,Agrawal-book}.
For the nonlocal case, because 
$\gamma$ depends linearly on $\widetilde{R}(k_{x})$, the instability property is sensitive to the response function. 
In the case of the response functions with a positive-definite spectrum~\cite{Wyller-pre-02}, such as the Gaussian one and the exponential-decay one, $\gamma>0$ determines that MI can only occur when $n_{2}>0$~\cite{Krolikowski-pre-01}. With the same reason, MI may happen both when $n_{2}>0$ and when $n_{2}<0$ in the case that the response functions have either positive or
negative spectral values, such as the rectangular response function~\cite{Krolikowski-pre-01} and the sine-oscillatory response function discussed in this paper.

\section{\label{sec-5}CONCLUTION}
We study the properties of the MI in the nonlocal optical Kerr medium with the sine-oscillatory response function in the framework of the nonlocal nonlinear Schr\"{o}dinger equation. We find that MI exists in both cases of negative Kerr coefficient $(n_{2}<0)$ and positive Kerr coefficient $(n_{2}>0)$. The MI in the case that $n_{2}<0$ can support the stability of the bright solitons and the instability of the dark solitons in the NLC with negative dielectric anisotropy. In addition, the largest MI gain fixedly occurs at the point of $|k_{x}|=1/w_{m}$ in the gain spectrum, and such a property is different from that in the other nonlocal media presented before.

We also discuss the physical mechanism of the MI generally in the framework of the FWM interaction. We find that the gain spectra of the MI depends on the competition between the phase mismatch $\Delta k$ and the (nonlinear) growth factor $\gamma$ in the FWM process. The necessary and sufficient condition for the MI is that the phase mismatch should be small enough such that $|\Delta k|<2|\gamma|$. The deduction is that  $\gamma>0$, that is
$n_{2}\widetilde{R}>0$, is a necessary condition for the MI existence, and the gain bands for the MI is determined by $0<|k_{x}|<2\sqrt{\gamma k}$. 
This conclusion can explain the absence of MI in the case of negative $n_{2}$ in the local Kerr media 
 and the nonlocal optical Kerr media with the positive-definite response spectral functions, and support the existence of MI in the both cases that $n_{2}>0$ and $n_{2}<0$ in the nonlocal optical Kerr media with the sine-oscillatory response and the rectangular response. 

\begin{acknowledgments}
This research was supported by the National Natural Science
Foundation of China, Grant No.~11474109.
\end{acknowledgments}

\appendix
\section{}
In the local optical Kerr media, the part of the third-order nonlinear polarization that influences the propagation of a linear polarized beam at the frequency $\omega$ is~\cite{book-Boyd1}
\begin{equation}\label{equ-a}
P^{\text{NL}}_{\text{lc}}(\omega)=3\epsilon_{0}\chi^{(3)}(\omega+\omega-\omega)\left|U\right|^{2}U,
\end{equation}
where $\chi^{(3)}$ is the third-order susceptibility. In order to relate the nonlinear polarization $P^{\text{NL}}_{\text{lc}}$ to the NRI $\Delta n$, we note that it is generally true~\cite{book-Boyd1}
\begin{equation}\label{equ-b}
\left(n_{0}+\Delta n\right)^{2}=1+\chi^{(1)}+3\chi^{(3)}\left|U\right|^{2},
\end{equation}
where $\chi^{(1)}$ is the linear susceptibility. By neglecting the quadratic term of $\Delta n$ $(\Delta n\ll n_{0})$ in Eq.~(\ref{equ-b}) and using the relation $n_{0}^{2}=1+\chi^{(1)}$, we get the expression that $\Delta n=3\chi^{(3)}\left|U\right|^{2}/2n_{0}.$ Thus the relation between the nonlinear polarization and the NRI in the local optical Kerr media is that $P^{\text{NL}}_{\text{lc}}=2n_{0}\epsilon_{0}\Delta nU$. Then by extending
the relation to the nonlocal case via Eq.~(\ref{equ-0}), the nonlinear polarization in the nonlocal optical Kerr
media is
\begin{equation}\label{equ-d}
P^{\text{NL}}_{\text{nonlc}}=2n_{0}\epsilon_{0}n_{2}U\int^{\infty}_{-\infty}R(x-x')|U(x',z)|^{2}dx'.
\end{equation}


\begin{thebibliography}{99}
\bibitem{footnote}
The substitution of Eq.~(\ref{equ-28}) into (\ref{equ-36a}) yields $|k_{x}^{2}/k-2\gamma|<2|\gamma|$. If $\gamma<0$, it is impossible that the sum of two positive numbers will be less than either of them. 
Therefore $\gamma>0$, i.e. $n_{2}\widetilde{R}>0$ [by Eq.~(\ref{equ-26b})], is necessary for MI. 
If $k_{x}^{2}/k-2\gamma\geq0$, the inequalities $k_{x}^{2}/k-2\gamma<2\gamma$ and  $k_{x}^{2}/k-2\gamma\geq0$ will give that $\sqrt{2\gamma k}\leq|k_{x}|<2\sqrt{\gamma k}$; on the other hand, the inequalities $k_{x}^{2}/k-2\gamma<0$ and $k_{x}^{2}/k-2\gamma>-2\gamma$ will yield the bands $0<|k_{x}|<\sqrt{2\gamma k}$. Thus the total spectral region for the MI is $0<|k_{x}|<2\sqrt{\gamma k}$.
\end{thebibliography}
\end{document}